# THE W AND Z BOSON SPIN OBSERVABLES AS MESSENGERS OF NEW PHYSICS AT LHC

**Jose Bernabeu**[1]
*Departamento de Física Teórica, Universidad de Valencia, and IFIC, Joint Centre Univ. Valencia-CSIC Burjassot, Valencia, Spain*
jose.bernabeu@uv.es

**Alejandro Segarra**
*Departamento de Física Teórica, Universidad de Valencia, and IFIC, Joint Centre Univ. Valencia-CSIC Burjassot, Valencia, Spain*
Alejandro.Segarra@uv.es

The successful LHC operation suggests going beyond the search of excess of events for the quest of new physics. We demonstrate that the eight multipole parameters describing the spin state of the W or Z bosons are able to disentangle their hidden production mechanism. They can be separately extracted from well defined angular asymmetries in the leptonic distribution of boson decays. The discriminating power of this analysis is well illustrated by: (i) polarised top quark decays, (ii) two body decay of heavy resonances, (iii) Drell-Yan production of Z plus jets, (iv) Z boson plus missing transverse energy.

[1]Speaker

## 1. Beyond "EXCESS OF EVENTS”

With the successful operation of LHC [1], accumulating a wealth of data in the ATLAS and CMS experiments at CM energies of 7, 8 and 13 TeV, and the expected increasing statistics in the coming years, precise measurements beyond simple event counts are mandatory. In Fig. 1 we present the LHC integrated luminosity by year in 2011, 2012, 2015 and 2016 along each year

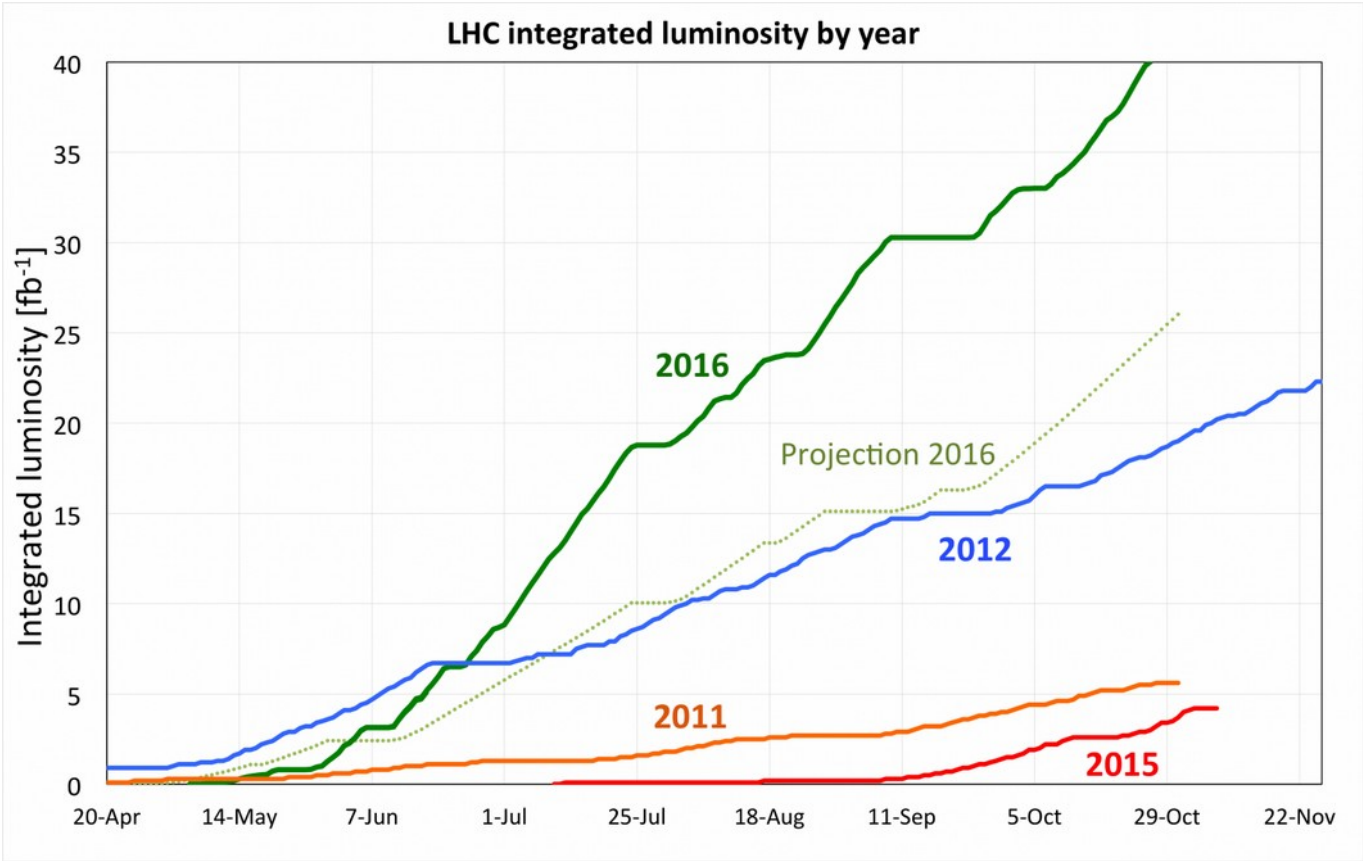

Fig. 1.- The LHC integrated luminosity by year operated at 7. 8 and 13 TeV. See [1].

Polarisation measurements are possible for particles with a lifetime leading to decay products within the detector from the angular distributions of these final states. The interest of these measurements is associated to their discriminating power between the Standard Model (SM) and New Physics scenarios for the production mechanism of the parent particles. W and Z bosons are particularly promising because, with spin 1 and massive, they are a natural vehicle at the electroweak energy scale and they offer the leptonic decays in the LHC detectors as superb polarisation analysers.

In Section 2 we discuss the formalism of the spin density matrix for describing the spin observables of the W, Z bosons in terms of the irreducible tensor operators with definite transformation properties under rotations. Their expectation values are the multipole parameters, polarisations and alignments, and we establish in Section 3 a biunivocal correspondence between them and definite angular asymmetries in the leptonic distribution of the Vector Boson decay. Section 4 illustrates the discriminating power of these W or Z boson spin observables for their different production mechanisms in polarised top quark decay, other heavy particle decays, Drell-Yan Z+jets and Z boson+MET processes. Our Conclusions and Outlook are summarized in Section 5.

## 2. SPIN DENSITY MATRIX

A quantum mixed state for massive spin 1 bosons has 8 independent Spin Observables. If we fix a coordinate system (x', y', z') in the boson rest frame, we can write the 3x3 spin density matrix as [2]

$$\rho = \frac{1}{3}\mathbb{1} + \frac{1}{2}\sum_{M=-1}^{1} \langle S_M \rangle^* S_M + \sum_{M=-2}^{2} \langle T_M \rangle^* T_M \,, \tag{1}$$

where the expectation values of the irreducible tensors are the multipole parameters, 3 for L=1 polarisations and 5 for L=2 alignments.

The angular distribution of the boson decay products is determined by ρ. We consider the leptonic decays. Using the helicity formalism [3], the amplitude for the decay of the boson with third spin component m giving leptons of helicities $\lambda_1$ and $\lambda_2$ is written as

$$\mathcal{M}_{m\lambda_1\lambda_2} = b_{\lambda_1\lambda_2} D^{1*}_{m\lambda}(\phi^*, \theta^*, 0)\,, \tag{2}$$

where $\lambda = \lambda_1 - \lambda_2$, and $D^j_{m'm}(\alpha, \beta, \gamma)$ are the Wigner D functions. For the W boson the left-handed interaction fixes a single helicity amplitude for each λ. For the Z boson we have two possible helicity combinations and the b's of Eq. (2) are proportional to the right- and left-handed couplings of the Z to the charged leptons. As a consequence, the terms in the angular distribution of leptons associated to the three Z polarisations are affected by a "polarisation analyser" given by

$$\eta_\ell = \frac{(g_L^\ell)^2 - (g_R^\ell)^2}{(g_L^\ell)^2 + (g_R^\ell)^2} = \frac{1 - 4s_W^2}{1 - 4s_W^2 + 8s_W^4}\,, \tag{3}$$

## 3. ANGULAR ASYMMETRIES versus BOSON MULTIPOLE PARAMETERS

The angular asymmetries introduced in [4, 5] for the measurement of the boson spin observables are able to separate out one by one the eight multipole parameters. This byunivocal correspondence is established as

$$\begin{aligned}
A_{\rm FB}^{x'} &= \frac{1}{\Gamma}\left[\Gamma(\cos\phi^* > 0) - \Gamma(\cos\phi^* < 0)\right] = -\frac{3}{4}\eta_\ell\langle S_1\rangle\,, \\
A_{\rm FB}^{y'} &= \frac{1}{\Gamma}\left[\Gamma(\sin\phi^* > 0) - \Gamma(\sin\phi^* < 0)\right] = -\frac{3}{4}\eta_\ell\langle S_2\rangle\,, \\
A_{\rm FB}^{z'} &= \frac{1}{\Gamma}\left[\Gamma(\cos\theta^* > 0) - \Gamma(\cos\theta^* < 0)\right] = -\frac{3}{4}\eta_\ell\langle S_3\rangle\,, \\
A_{\rm EC}^{z'} &= \frac{1}{\Gamma}\left[\Gamma(|\cos\theta^*| > \frac{1}{2}) - \Gamma(|\cos\theta^*| < \frac{1}{2})\right] = \frac{3}{8}\sqrt{\frac{3}{2}}\langle T_0\rangle\,, \\
A_{\rm FB}^{x',z'} &= \frac{1}{\Gamma}\left[\Gamma(\cos\phi^*\cos\theta^* > 0) - \Gamma(\cos\phi^*\cos\theta^* < 0)\right] = -\frac{2}{\pi}\langle A_1\rangle\,, \\
A_{\rm FB}^{y',z'} &= \frac{1}{\Gamma}\left[\Gamma(\sin\phi^*\cos\theta^* > 0) - \Gamma(\sin\phi^*\cos\theta^* < 0)\right] = -\frac{2}{\pi}\langle A_2\rangle\,, \\
A_\phi^1 &= \frac{1}{\Gamma}\left[\Gamma(\cos 2\phi^* > 0) - \Gamma(\cos 2\phi^* < 0)\right] = \frac{2}{\pi}\langle B_1\rangle\,, \\
A_\phi^2 &= \frac{1}{\Gamma}\left[\Gamma(\sin 2\phi^* > 0) - \Gamma(\sin 2\phi^* < 0)\right] = \frac{2}{\pi}\langle B_2\rangle\,.
\end{aligned} \tag{4}$$

and it is picturically given in Fig. 2 for the three polarisations and the five alignments

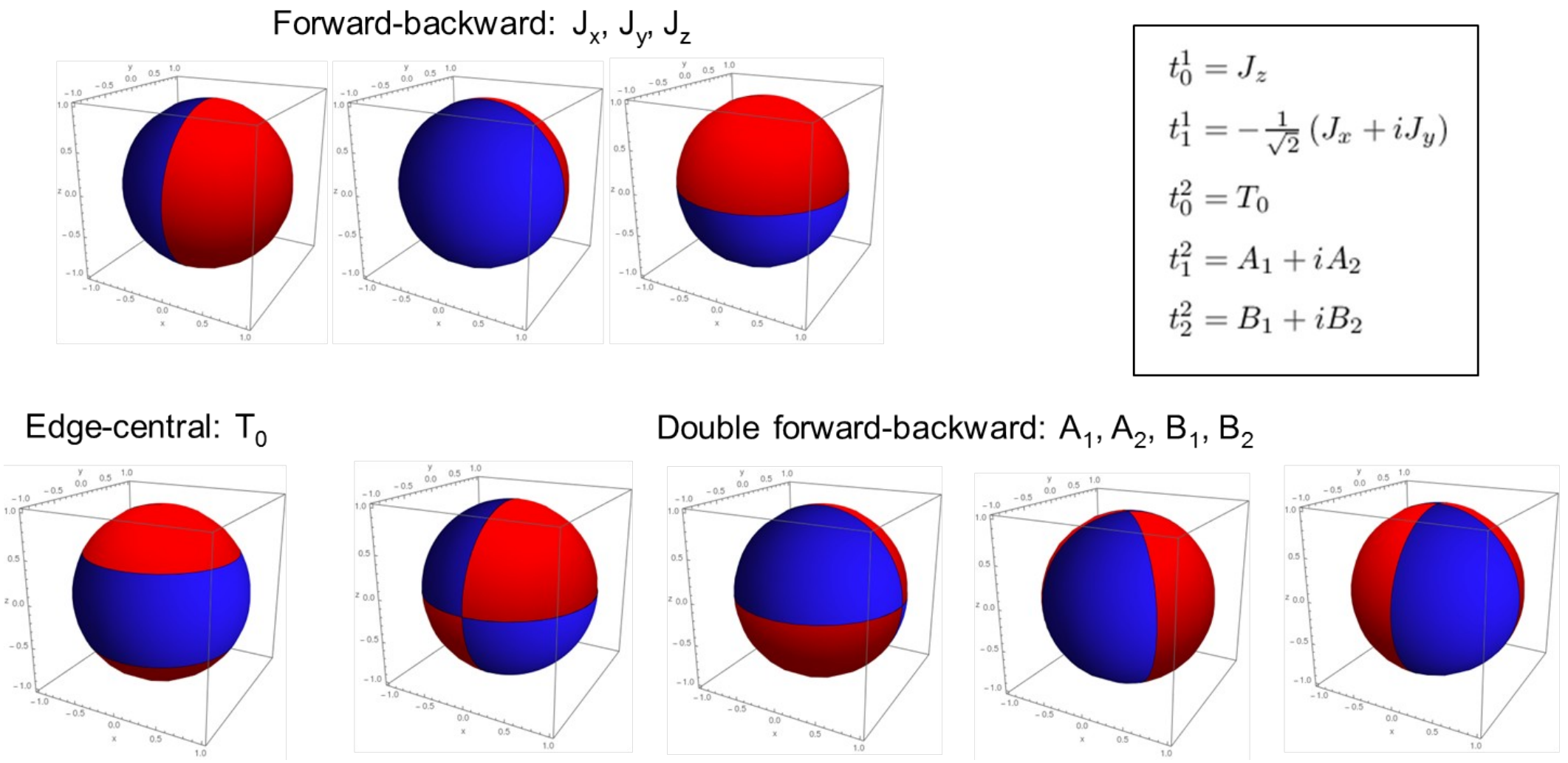

Fig. 2.- The eight angular asymmetries of the lepton distribution for separating out the eight spin observables of the parent boson.

## 4. DISCRIMINATING POWER

In this Section we consider different processes with model-dependent production mechanisms for the bosons and calculate the corresponding spin observables.

### 4.1 Polarised Top Quark Decay

In general, the top quark decay amplitude can be calculated with a vertex that incorporates left- and right-handed vebtor (chirality conserving) and tensor (chirality changing) couplings.

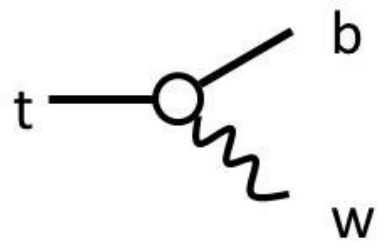

Since $\rho_{\pm1,\mp1} = 0$ for the W boson in this process, we have that the two B's expectation values associated to $T_2^2$ vanish $\langle B_1 \rangle = \langle B_2 \rangle = 0$. The other six multipole parameters are given in Table 1 for the Standard Model (SM) and for a model which has, in addition, a right-handed tensor coupling $g_R$. This additional dipole interaction, with an imaginary coupling has special interest because it is able to interfere with the SM coupling, generating a polarisation component normal to the decay plane and it could be a novel source of CP Violation

| | $\langle S_1 \rangle$ | $\langle S_2 \rangle$ | $\langle S_3 \rangle$ | $\langle T_0 \rangle$ | $\langle A_1 \rangle$ | $\langle A_2 \rangle$ |
|---|---|---|---|---|---|---|
| SM | 0.510 | 0 | -0.302 | -0.445 | 0.255 | 0 |
| $g_R = 0.03$ | 0.500 | 0 | -0.278 | -0.472 | 0.249 | 0 |
| $g_R = 0.10\,i$ | 0.507 | -0.084 | -0.284 | -0.434 | 0.253 | -0.042 |

Table 1.- Spin Observables for the W boson in the decay of polarized top quarks.

All the transverse observables $<S_{1,2}>$ and $<A_{1,2}>$ need a Top Quark Polarisation in order to define an azimuthal plane, which is generated for single top production by weak interactions. In addition, for ensuring that $<S_2>$ and/or $<A_2>$ are bona fide signals of CP Violation one should compare the measurements for top and antitop decays. Their sensitivity to an Im $g_R$ is very welcome.

The ATLAS Collaboration has measured [6] the angular coefficients of the lepton distribution in polarised quark decays, with the results shown in Fig. 3

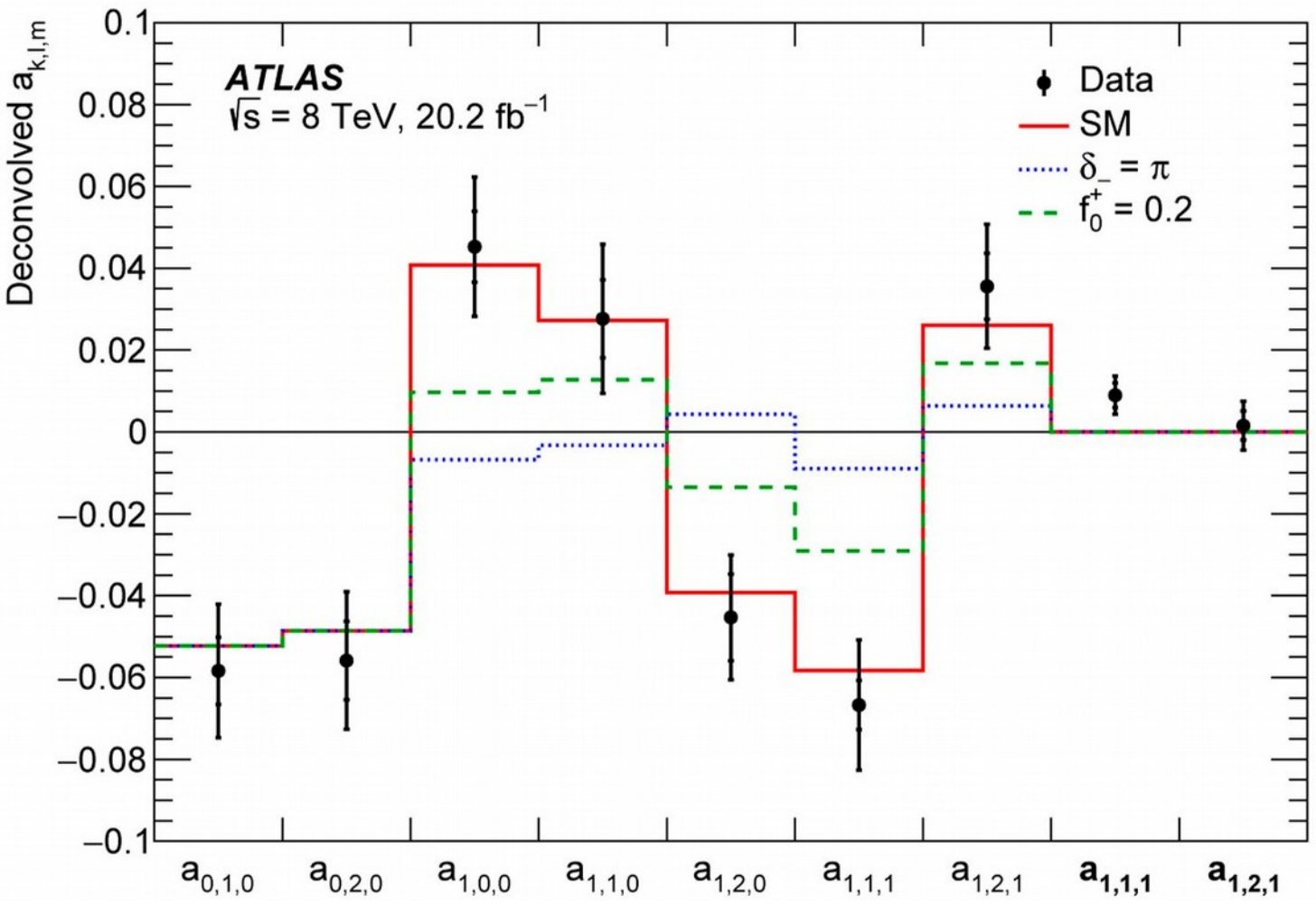

Figure 3. Deconvolved angular coefficients from data using the migration matrix from the SM simulation. Data are shown as black points with statistical uncertainties (inner error bar) and statistical and systematic uncertainties added in quadrature (outer error bar), while SM prediction is shown as a red line. In addition, two new physics scenarios are also shown as a dotted blue line and dashed green line, respectively.

## 4.2 Heavy particle decays

Let us consider that a W or Z boson is produced in the two body decay of some spin j particle A yielding also a spin j' particle B as decay product. The case $j=1/2 \rightarrow Z + j'=1/2$ would have a similar analysis to that discussed in 4.1 for $t \rightarrow W + b$.

A most interesting case would be for $j = 0 \rightarrow Z + j' = 0$. Under these restrictions, the Z has to be fully longitudinal with $\lambda = 0$. This is equivalent to say that we have a pure P-wave L=1 decay. You may think in a heavy scalar decaying to Z plus the standard Higgs, like the SUSY pseudoscalar $A(0^-) \rightarrow Z + h(0^+)$. The only non-vanishing matrix element of the spin density matrix is $\rho_{00}= 1$, leading to an alignment $<T_0> = -2/\sqrt{6}$

Another particular case of high interest is $j = 0 \rightarrow Z + j' =1$, which corresponds to Di-Boson Resonances. With these spins, the spin density matrix of the Z boson is diagonal, leading to the two longitudinal polarisation and alignment

$$\langle S_3 \rangle = \left[ \left| a_{1,1} \right|^2 - \left| a_{-1,-1} \right|^2 \right] / N$$
$$\langle T_0 \rangle = \frac{1}{\sqrt{6}} \left[ 1 - 3 |a_{00}|^2 / N \right] \qquad (5)$$

where the a's are the corresponding diagonal helicity amplitudes of the decay.

### 4.3 Drell-Yan Z + jets production

The angular distribution of leptonic Z decays has been investigated by the CDF [7], CMS [8] and ATLAS [9] Collaborations for Drell-Yan Z production in hadron collisions. The leading SM amplitudes are obtained from the production mechanism

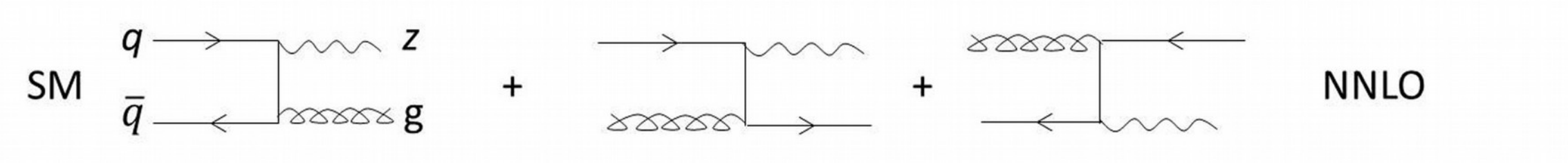

With our spin density matrix analysis, we interpret the measured coefficints in terms of the Z boson spin observables as given by

$$\langle S_3 \rangle = \left[ \left| a_{1,1} \right|^2 - \left| a_{-1,-1} \right|^2 \right] / N$$

$$\langle T_0 \rangle = \frac{1}{\sqrt{6}} \left[ 1 - 3 |a_{00}|^2 / N \right] \qquad (6)$$

The measurement of these coefficients is made differentially, as function of the transverse momentum and rapidity of the Z boson. In Fig. 4 we present the polar and azimuthal angle distributions as measured by CMS at 8 TeV in the indicated region of transverse momentum

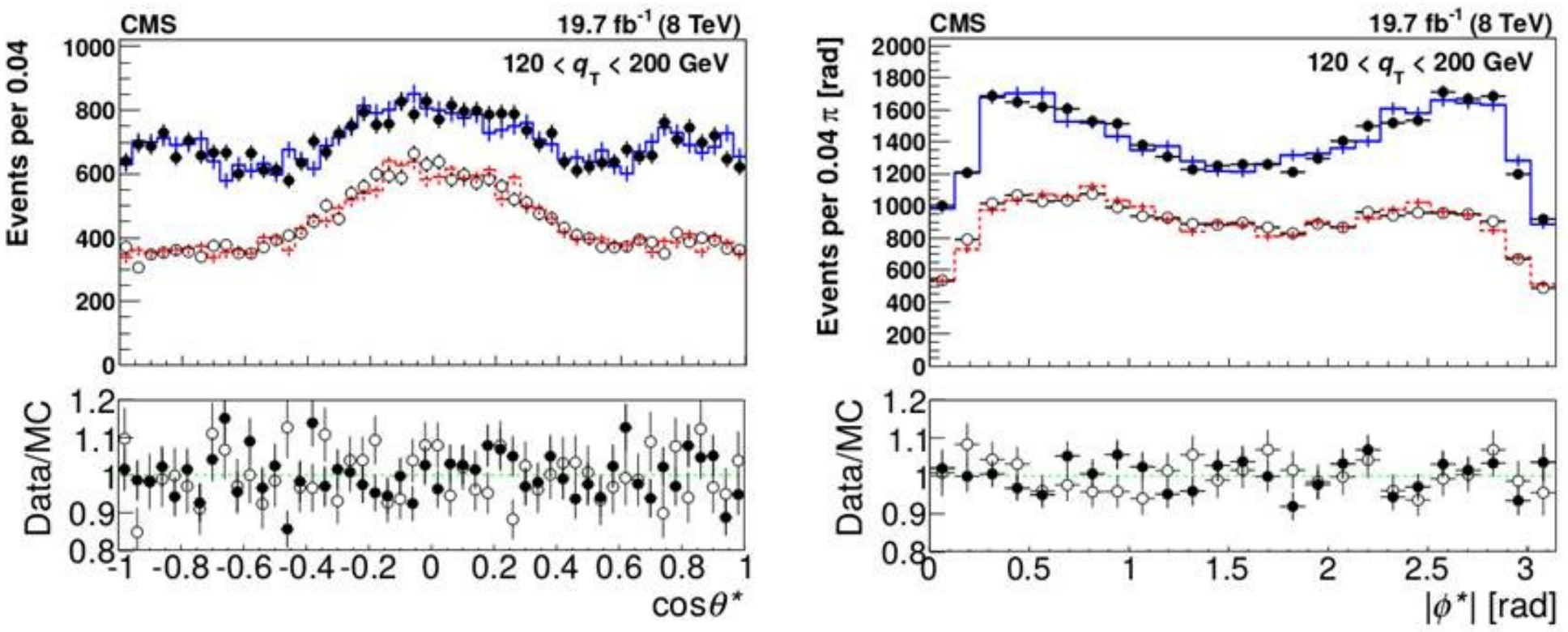

Fig. 4.- Polar and azimuthal angular distributions of leptons for Drell-Yan Z production as measured by CMS.

whereas in Fig. 5 we present the most recent measurement by the ATLAS Collaboration exhibiting a noticeable tension in $A_2$ with respect to next-to-next-to-leading order SM predictions at large transverse momentum of the Z boson

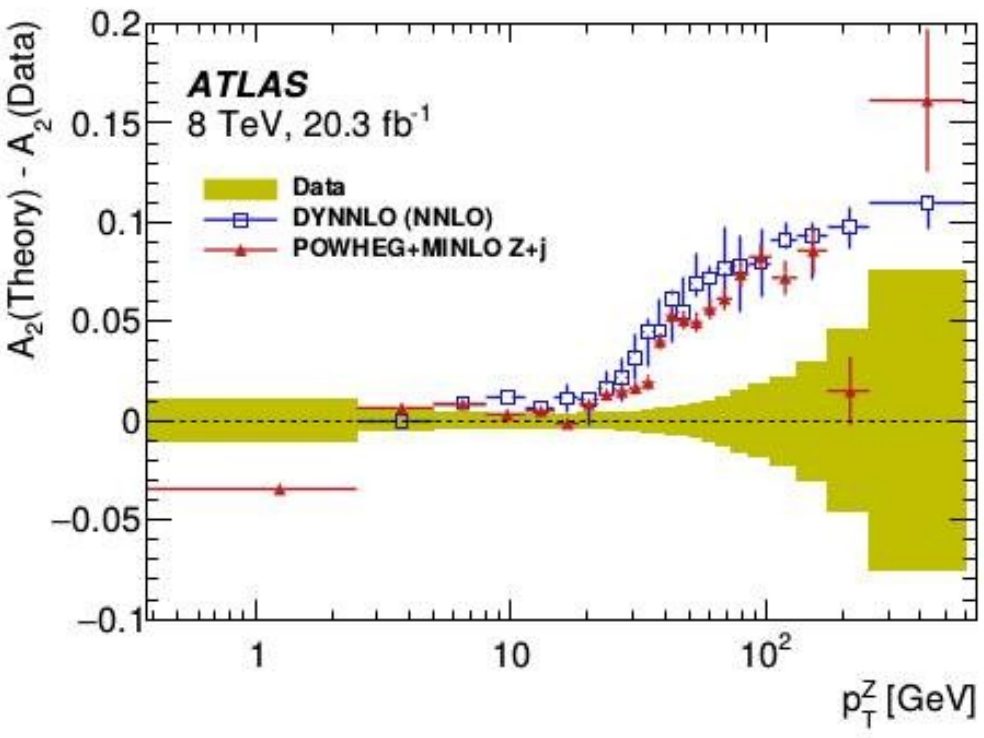

Fig. 5.- The Difference between $A_2$(Theory)-$A_2$(Data) for the ATLAS measurement as function of the transverse momentum.

This coefficient $A_2$ corresponds in our approach to the rank-two transverse alignment $\langle B_1 \rangle$ with $M = \pm 2$.

### 4.4 Z boson plus MET production

This is considered to be a priviledged process for the search of SUSY signals at LHC and for Dark Matter production. The proposal is the measurement of the angular distribution of l's in $Z \to l\, l$, identifying the angular coefficients as function of the final missing transverse energy (MET).

The leading order SM production mechanisms are ZZ production with $ZZ \to l\, l\, \upsilon\, \upsilon$ and ZW production with the additional charged lepton undetected

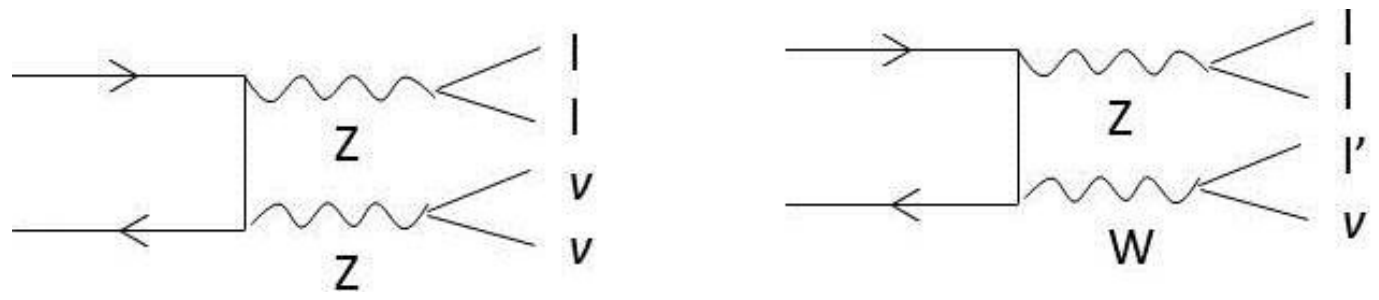

A simulation at 13 TeV pp collisions has been performed [5] with the Z direction as unique reference. As a consequence, one has access to the two longitudinal observables $\langle S_3 \rangle$ and $\langle T_0 \rangle$ only.

They show a very interesting dependence in the SM on the MET cut above 100 GeV, as shown in Fig. 6

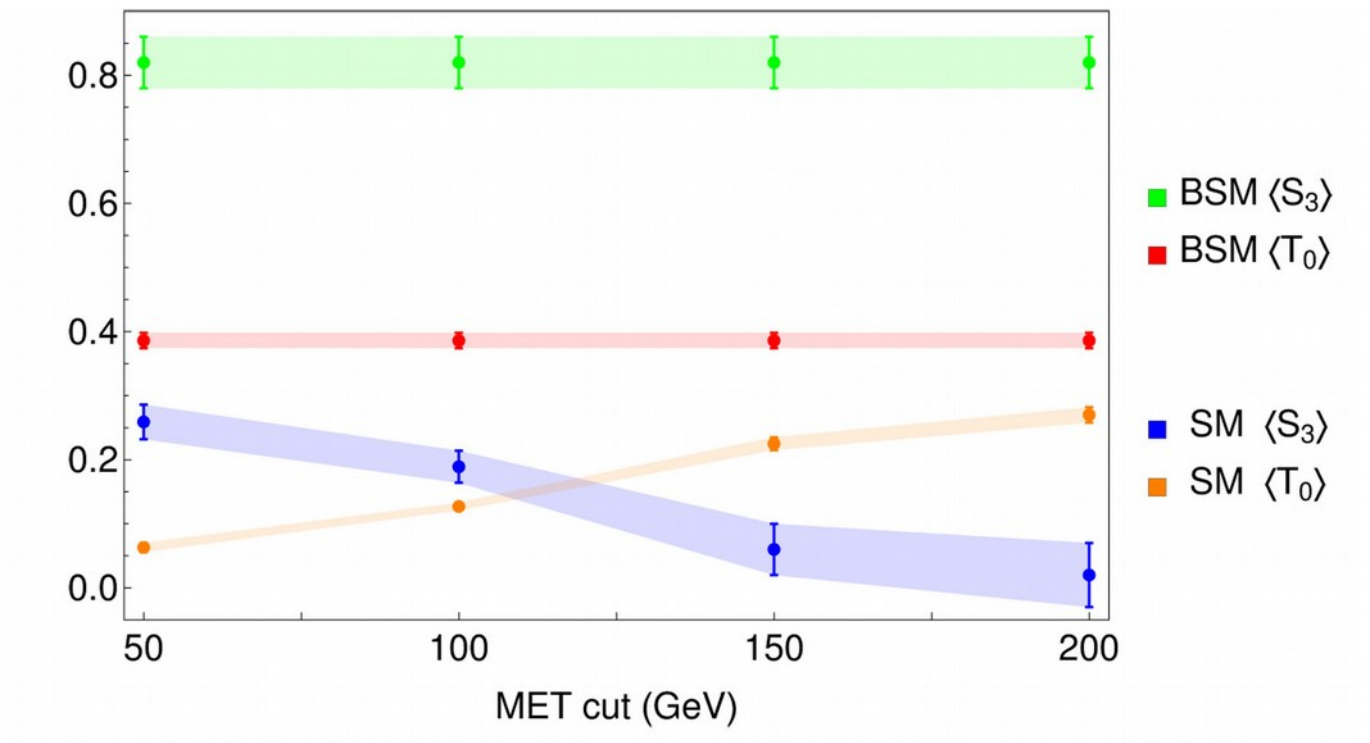

Fig. 6. SM predictions for $\langle S_3 \rangle$ and $\langle T_0 \rangle$ for Z + MET final states, as a function of the lower cut on MET. Above them, the same result from the BSM model described in the text.

To illustrate the power of the declared strategy, we compare with the expected values of these spin observables in a SUSY dark matter model with the gravitino as lightest supersymmetric particle (LSP) and the lightest neutralino as next-to-LSP. Taking a direct electroweak production of the neutralino pair, as shown in the diagrams the two spin observables are independent of the SM cut and clearly different in behaviour and value to the SM results. For a Z production from the decay $j=1/2 \to Z + j' = 3/2$ the predicted alignment $\langle T_0 \rangle = 1/\sqrt{6}$ is obtained, in agreement of the result obtained in our simulation. These results are also plotted in Fig. 6.

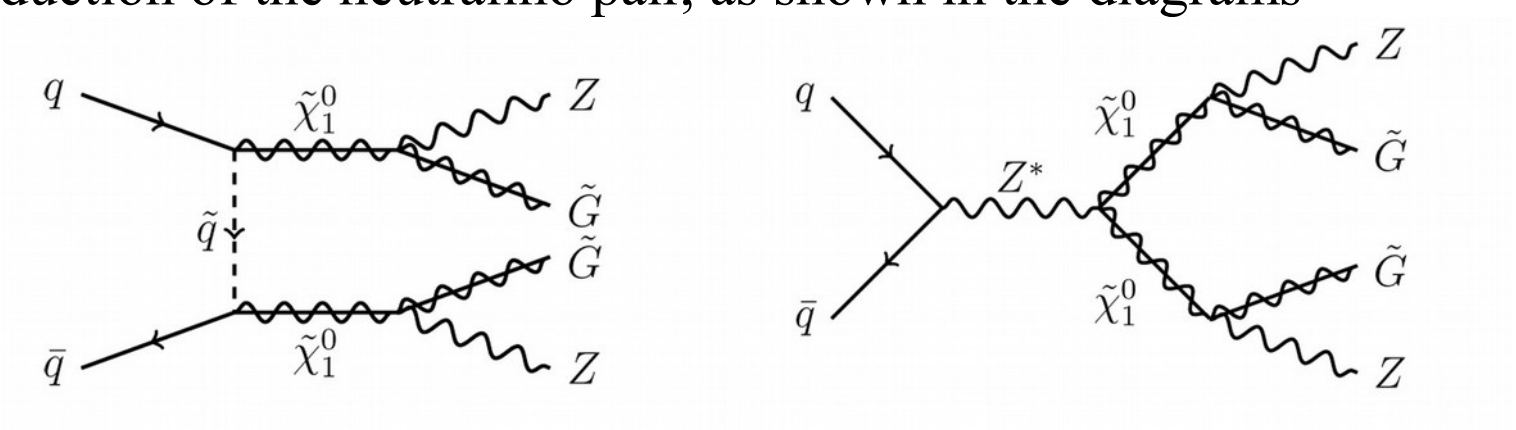

## 5. CONCLUSIONS AND OUTLOOK

With the wealth of LHC collision data, the separate measurement of the eight W and Z boson spin observables is becoming feasible. The methodology to be followed is well defined by definite asymmetries in the angular distribution of leptons in the boson decay.

The reason for undertaking such a programme is very illuminating: these observables, which are independent of PDF's, offer a discriminating power of the hidden production mechanism of the vector boson, acting as messengers of either the SM or BSM New Physics scenarios. As illustrations,

- The W boson spin properties in $t \rightarrow W b$ decay distinguish the SM from an additional dipole.
- The two body decay of heavy resonances involving W or Z boson shows that different spin assignments lead to specific zeros and values of the boson spin observables.
- For Drell-Yan Z production, some tension in the identified transverse $M=\pm 2$ Alignment al large transverse momentum is apparent.
- For the Z boson plus MET production there is a very interesting rapid variation of the longitudinal polarisation $\langle S_3 \rangle$ and alignment $\langle T_0 \rangle$ in the SM above 100 GeV of MET, characteristic of SM. Different and constant values of these observables are obtained in a model for the neutralino decay to Z plus the gravitino.

To summarize, in the search for New Physics there is an invaluable interesting programme of measurements of the Boson Spin Observables when reconstructing the W or Z bosons from their leptonic decays in different processes.

### ACKNOWLEDGEMENTS

This research has been supported by MINECO Project FPA 2014-54459-P, Generalitat Valenciana Project GV PROMETEO II 2013-017 and Severo Ochoa Excellence Centre Project SEV-2014-0398. A.S. acknowledges the MECD support through the FPU 14/04678 grant.